\documentclass[prl,twocolumn,nofootinbib]{revtex4}
\usepackage{graphicx}
\usepackage{latexsym}
\def\cm{{\rm cm}}
\def\be{\begin{equation}}
\def\ee{\end{equation}}
\def\bea{\begin{eqnarray}}
\def\eea{\end{eqnarray}}

\begin{document}
\title{Searching for CPT Violation with Cosmic Microwave Background Data from WMAP and BOOMERANG}

\author{Bo Feng${}^{a,b}$}

\author{Mingzhe Li${}^c$}

\author{Jun-Qing Xia${}^d$}

\author{Xuelei Chen${}^a$}

\author{Xinmin Zhang${}^d$}

\affiliation{ ${}^a$ National Astronomical Observatories, Chinese
Academy of Sciences, Beijing 100012, P. R. China}

\affiliation{ ${}^b$ Research Center for the Early Universe(RESCEU),
Graduate School of Science, The University of Tokyo, Tokyo 113-0033,
Japan}

\affiliation{$^c$ Institut f\"{u}r Theoretische Physik,
Philosophenweg 16, 69120 Heidelberg, Germany}

\affiliation{${}^d$Institute of High Energy Physics, Chinese Academy
of Science, P.O. Box 918-4, Beijing 100049, P. R. China}


\begin{abstract}
We search for signatures of Lorentz and $CPT$ violations in the
cosmic microwave background (CMB) temperature and polarization
anisotropies by using the WMAP and the 2003 flight of BOOMERANG
(B03) data. We note that if the Lorentz and $CPT$ symmetries are
broken by a Chern-Simons term in the effective lagrangian, which
couples the dual electromagnetic field strength tensor to an
external four-vector, the polarization vectors of propagating CMB
photons will get rotated. Using the WMAP data alone, one could put
an interesting constraint on the size of such a term. Combined with
the B03 data, we found that a nonzero rotation angle of the photons
is mildly favored: $\Delta \alpha= -6.0^{+4.0}_{-4.0}$
$^{+3.9}_{-3.7}$ deg (1, 2 $\sigma$ ).

\end{abstract}

\maketitle


After decades of pursuance and many advances in both the theoretical
and the observational fronts of cosmological research, a ``standard
model'' of structure formation has been established. It is now
possible, with the unprecedented precision of the cosmological
observations\cite{wmap,sloanx}, to have robust tests and effective
distinctions of the many theoretical models of new physics.

A possible signature of new physics is the $CPT$ violation. In the
standard model of particle physics $CPT$ is an exact symmetry. The
detection of $CPT$ violation will reveal new physics beyond the
standard model. There have been a number of high precision
experimental tests on $CPT$ conservation in the laboratory
\cite{kostelecky}. Now cosmology provides another way to test this
important symmetry. Also, breaking of the $CPT$ symmetry may have
played an active role in cosmological evolution. For example,
$CPT$-violating interactions in the baryons and leptons provide a
baryogenesis mechanism where the baryon number asymmetry is produced
in thermal equilibrium
\cite{cohenkaplan,kostelecky2,li,steinhardt,hongli,alberghi}.

In this paper we study the cosmological $CPT$ violation in the photon sector.
 Phenomenologically, we introduce a Chern-Simons term in the effective
 lagrangian of the form \cite{jackiw,field}
 \be\label{CFJCS}
{\mathcal L}_{int} \sim p_{\mu} A_{\nu}\widetilde{F}^{\mu\nu}~, \ee
where $p_{\mu}$
 is a four vector and
 $\widetilde{F}^{\mu\nu}=(1/2)\epsilon^{\mu\nu\rho\sigma}
F_{\rho\sigma}$ is the dual of the electromagnetic tensor. The
action of (\ref{CFJCS}) is gauge invariant if $p_{\mu}$ is a
constant and homogeneous vector or the gradient of a scalar. It
violates Lorentz and $CPT$ symmetries if the background value of
$p_{\mu}$ is non-zero. In the scenario of quintessential
baryo/leptogenesis the four vector $p_\mu$ is in the form of the
derivative of the quintessence\cite{quint} scalar,
$\partial_{\mu}\phi$. During the evolution of quintessence, the time
component of $\partial_{\mu}\phi$ does not vanish, which causes
$CPT$ violation. In the scenario of gravitational baryo/leptogenesis
\cite{steinhardt,hongli}, $p_\mu$ is the gradient of a function of
Ricci scalar $R$ \cite{lehnert}.

The interaction in (\ref{CFJCS}) violates also the $P$ and $CP$
symmetries, as long as $p_{0}$ dose not vanish (e.g., in the case
$p_{\mu}$ is the gradient of a time dependent scalar field)
\cite{klinkhamer}. This term can lead to a rotation of the
polarization vector of electromagnetic waves when they are
propagating over cosmological distances \cite{jackiw}. This effect
is known as ``cosmological birefringence''. The change in the
position angle of the polarization plane $\Delta\alpha$, which is
obtained by observing polarized radiation from distant radio
galaxies and quasars, provides a sensitive measure of the strength
of the cosmological birefringence, and this has been used to
constrain modified electrodynamics \cite{jackiw,field,CF97,carroll}.

In the present paper we use the current CMB polarization data to
measure this type of Lorentz- and $CPT$-violating term  {\it for the
first time}. Our results show that the current CMB data provide an
interesting indication for a nonzero $p_{\mu}$.

The Stokes parameters Q and U of the CMB polarization can be
decomposed into a gradient-like (G) and a curl-like (C) components
\cite{HW97}. If parity is not violated in the
temperature/polarization distribution, the TC and GC cross
correlation power spectra vanish due to the intrinsic properties of
the tensor spherical harmonics. With the presence of cosmological
birefringence, the polarization vector of each photon is rotated by
an angle $\Delta \alpha$, and one would observe non-zero TC and GC
correlations, even if they are zero at the last scattering surface.
Denoting the rotated quantities with a prime, one gets
\cite{kamionkowski}
\begin{equation}
\label{TC} C_l'^{TC}= C_l^{TG}\sin 2 \Delta
\alpha~
\end{equation}
and\cite{FLLZ} \begin{equation}
\label{GC} C_l'^{GC}=
\frac{1}{2}(C_l^{GG}- C_l^{CC})\sin 4 \Delta \alpha ~.
\end{equation}
On the other hand, the original TG, GG and CC spectra are also
modified:
\begin{eqnarray}
\label{TG}
C_l'^{TG} &=& C_l^{TG}\cos 2 \Delta \alpha ~,\\
\label{GG}
C_l'^{GG} &=& C_l^{GG}\cos^2 2 \Delta \alpha + C_l^{CC}\sin^2 2 \Delta \alpha~,\\
\label{CC}
C_l'^{CC} &=& C_l^{CC}\cos^2 2 \Delta \alpha + C_l^{GG}\sin^2 2\Delta \alpha~ ~.
 \end{eqnarray}

From the above discussion, we see that even with only the TG cross
correlation power spectrum (and the TT autocorrelation power
spectrum),
the Lorentz- and $CPT$-violating term can still be
measured. Of course, direct measurements of the TC and GC power
spectra would allow more stringent constraints. Indeed, the GC
spectrum will be the most sensitive probe of such Lorentz- and
$CPT$-violating term \cite{FLLZ}, this is because the GC power
spectrum is generated by the rotation of the GG power spectrum,
which is a more sensitive probe of the primordial fluctuation than
the TT and TG spectra \cite{kpola}.

To break possible degeneracy between this term and the variation of
other parameters, we make a global fit to the CMB data with the
publicly available Markov Chain Monte Carlo package
\texttt{cosmomc}\cite{Lewis:2002ah,IEMCMC}, which has been modified
to allow the rotation of the power spectra discussed above, with a
new free parameter $\Delta \alpha$. We assume purely adiabatic
initial conditions, and impose the flatness condition motivated by
inflation. The following set of 8 cosmological parameters are
sampled: the Hubble constant $h$, the physical baryon and cold dark
matter densities, $\omega_b=\Omega_bh^2$ and $\omega_c=\Omega_ch^2$,
the ratio of the sound horizon to the angular diameter distance at
decoupling, $\Theta_s$, the scalar spectral index and the overall
normalization of the spectrum, $n_s$ and $A_s$, the tensor to scalar
ratio of the primordial spectrum $r$, and, finally, the optical
depth to reionization, $\tau_r$. We have imposed the Guassian Hubble
Space Telescope prior\cite{freedman}: $h=0.72\pm 0.08$  and also a
weak big-bang nucleosynthesis prior\cite{bbn}: $\Omega_b h^2 = 0.022
\pm 0.002$ (1
$\sigma$). 
For the other parameters we have adopted flat priors, and in the
computation of the CMB spectra we have considered lensing
contributions.

When the first version of this paper was being prepared, the
available temperature and polarization power spectra included the
results from the first-year observation of WMAP
\cite{wmap,hinshaw,kogut,Verde03}, and those from the January 2003
Antarctic flight of BOOMERANG (Hereafter B03)\cite{B03,B03EE,TCGC}.
The WMAP three year data (WMAP3) has since been released, with
significant improvements on the estimate of the TE power at small
$\ell$ \cite{page}. We have repeated our analysis with the new WMAP
three year data.

\begin{figure}[htbp]
\begin{center}
\includegraphics[scale=0.55]{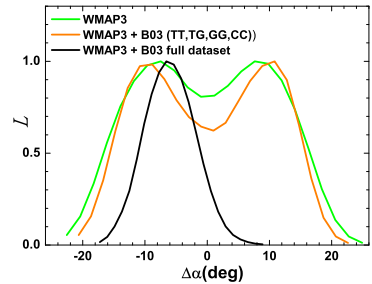}
\caption{ (color online). One dimensional constraints on the
rotation angle $\Delta \alpha$ from WMAP data alone (Green or light
gray line), WMAP and the 2003 flight of BOOMERANG B03 TT, TG, GG and
CC(Orange or gray line), and from WMAP and the full B03 observations
(TT, TG, GG, CC, TC, GC) (Black line). \label{fig:fig1}}
\end{center}
\end{figure}

In Fig.~\ref{fig:fig1} we plot our one dimensional constraints on
$\Delta \alpha$ from the WMAP data alone, and from the combined WMAP
and B03 data. We have assumed that the cosmic rotation is not too
large and imposed a flat prior $-\pi/2\le \Delta \alpha \le \pi/2$.
The CMB temperature power spectrum remains unchanged with the
rotation while the TG power spectrum gets modified, as given by Eq.
(\ref{TG}).

Using the data from WMAP alone, for both the first and three year
data set, we obtain a null detection within the error limits. For
WMAP3 the 1, 2 $\sigma$ constraints are $\Delta \alpha=
0.0^{+11.6}_{-11.7}$ $^{+5.9}_{-5.9}$ deg. The uncertainty is
considerable, as the error bars of the WMAP TG data are relatively
large, and TG data are not very sensitive probe. In the likelihood
of Fig.~\ref{fig:fig1} we have gained double peaks, which can be
easily understood from Eq.(\ref{TG}) due to the symmetry around
$\Delta \alpha = 0$.

With the inclusion of the B03 data, the measurement could be
improved dramatically. In a first step we also consider the indirect
measurements only by including the B03 TT, TG, GG and CC data. We
find the constraint on $\Delta \alpha$ becomes a bit more stringent
compared with WMAP only, a nonzero $\Delta \alpha$ is slightly
favored and the double peaks are still present. When the B03 TC and
GC data are also included the degeneracy around $\Delta \alpha = 0$
is broken.
We get the 1, 2 $\sigma$ constraints to be $\Delta \alpha=
-6.0^{+4.0}_{-4.0}$ $^{+3.9}_{-3.7}$ deg with WMAP3 and the B03 full
data set.

\begin{figure}[htbp]
\begin{center}
\includegraphics[scale=0.55]{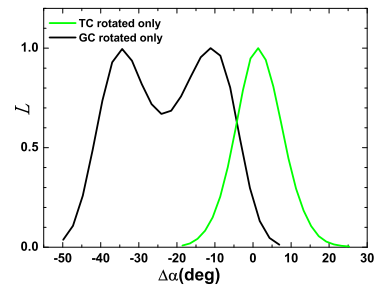}
\caption{ (color online). One dimensional constraints on the
rotation angle $\Delta \alpha$ from WMAP and the B03 observations,
assuming only CMB TC is rotated to be nonzero by $\Delta
\alpha$(Green or gray line) and only GC rotated(Black line).
\label{fig:fig2}}
\end{center}
\end{figure}

The covariance matrices of the B03 TC and GC data are correlated. In
order to find out what role the TC and GC data play in our fitting
respectively, we have made fits with, in one case, only the TC
spectrum rotated as Eq. (\ref{TC}), and in the other case only the
GC spectrum rotated. To make the comparison clear and avoid the
problem of convergence we set the flat prior $-1.2 \le \Delta \alpha
\le 0.8$. In Fig.~\ref{fig:fig2}, we plot the resulting  one
dimensional constraints. In neither case is the likelihood symmetric
at $\Delta \alpha = 0$. In the TC rotated only case, the symmetric
points are around $\pm \pi/4$, as we can see from Eq.(\ref{TC}).
Such a symmetry is lost for this narrower prior, but in our global
fittings (Fig.~\ref{fig:fig1}) we have allowed a larger range of
$\Delta \alpha$. We find from Fig.~\ref{fig:fig2} the TC data are
very weak in breaking the degeneracy around $\Delta \alpha = 0$,
while for GC the rotation is more eminent, where the likelihood in
Fig.~\ref{fig:fig2} is centered around $\Delta \alpha = - \pi/8$. In
this fit $\Delta \alpha = 0$ has an excess of $\Delta \chi^2 = 4$,
which is disfavored compared with the best fit case.

\begin{figure}[htbp]
\begin{center}
\includegraphics[scale=0.3]{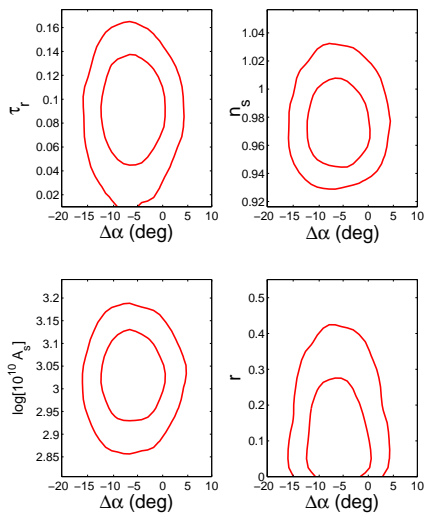}
\caption{ (color online). Joint 2-dimensional posterior probability
contour plots in the  $\Delta \alpha - \tau_r$ (top left), $\Delta
\alpha - n_S$ (top right), $\Delta \alpha - \log [10^{10} A_S]$
(bottom left) and $\Delta \alpha - r$ (bottom right), showing the
$68\%$ and $95\%$ contours from the WMAP + B03 constraints.
\label{fig:fig3}}
\end{center}
\end{figure}

The effect of the polarization rotation is degenerate with variation
on the amplitude of the primordial spectrum and the tensor to scalar
ratio. These parameters are also degenerate with the optical
depth of reionization. In
Fig.~\ref{fig:fig3} we plot the joint two-dimensional posterior
probability contours of $\Delta \alpha$ with $\tau_r$, $n_S$, $A_S$
and $r$. More precise measurements on these four parameters
will help to break the
degeneracy on the constraints of cosmic Lorentz and $CPT$ violations
discussed here. We have also made fits with a running spectrum index,
but found that it does not affect the above results significantly.
The inclusion of the matter power spectrum obtained from
large scale structure measurements also does not change
our constraints on $\Delta \alpha$ significantly.

Previously, the cosmological birefringence effect has been constrained
by looking for correlations between the elongation axes and polarization vectors
of distant radio galaxies and quasars. The most recent searches yield
null results, with an error on $\Delta \alpha$ at the order of
$1^{\circ}$ level \cite{carroll,field,CF97}. The typical redshifts of
the sources in these searches are of order of unity. Conceivably,
for the much greater redshift range between the last
scattering surface and the present-day observer, the
cumulative effect of cosmological birefringence could be
stronger.

It was claimed that the individual B03 CC and CG data are consistent
with zero \cite{B03EE}, however, we found that a negative rotation
angle is preferred in our combined analysis. It is noteworthy that
our result relies mainly on the fact that at $\l\sim 350$, GG power
of B03 is positive, CC power is (slightly) positive and GC power is
(slightly) negative. Given Eq.(\ref{GC}), the GC power spectrum
helps to increase the statistical significance on nonzero $\Delta
\alpha$. At present, the only publicly available (polarization) data
are the three-year WMAP and the data from a 200-hour flight of the
BOOMERANG balloon. In the coming few years, the quantity and quality
of the CMB polarization data are likely to be improved rapidly, with
the ongoing WMAP observations and many balloon experiments like the
BOOMERANG. These would allow better measurements of $\Delta\alpha$.

While nonzero TG and CG power can also be induced by Faraday
rotations\cite{Scannapieco:1997mt} and higher dimensional Lorentz
and $CPT$ violating operators\cite{Gambini:1998it}, these are often
frequency-dependent, while the effects described here are
not\cite{Pogosian:2001np}. This provides, at least in principle, a
way to distinguish between these different effects. The Faraday
rotation induced by magnetic field is given by
\begin{equation}
\frac{\Delta \alpha}{\rm rad} =8.1\times 10^{5}
\left(\frac{\lambda}{\rm m}\right)^2
\int_{0}^L \left(\frac{B_{\parallel}}{\rm Gs}\right)
\left(\frac{n_e}{\cm^{-3}}\right) \frac{dL}{\rm pc}.
\end{equation}
If we assume that reionization occurs at $z<20$, then for a global
intergalactic magnetic field of $10^{-9}$ Gs, at the frequency of
145 GHz where BOOMERANG operates, the Faraday rotation is only of
the order of $10^{-3}$ deg, which is much smaller than the range of
$\alpha$ uncertainty distribution and hence insignificant. The
apparent rotation might also be due to contamination from foreground
emission. In some attempts to obtain CMB temperature and
polarization spectra, including those of WMAP, foreground-removing
procedure has been applied. For the BOOMERANG experiment, which
operates at relatively high frequency, it is believed that the
primary CMB polarization signal is dominant, and the contribution of
the polarized galactic synchrotron foreground is small \cite{B03EE},
but at present a small contamination can not be ruled out
completely. Future multi-wavelength polarization observations would
help distinguish this possibility.

We could not yet conclude that $CPT$ is definitely violated if a
non-zero $\Delta\alpha$ is detected. However, if such a detection is
confirmed, it would certainly raise the possibility of a
Lorentz-violating term like that given in Eq.(\ref{CFJCS}), or
others of the similar form. For example, a term of this form could
be due to the interaction between dark energy and the
electromagnetic sector, if we take $p_{\mu}$ as $\partial_{\mu}\phi$
with $\phi$ being quintessence field. Thus, the results we obtained
can be used to put additional constraints on the behaviors of
dynamical dark energy between the redshift range $z\sim 1$ to $z
\sim 1000$.

A Lorentz violation also implies the violation of the equivalence
principle. In our case where only a small violation is present, the
group velocity of light remains unchanged, and the weak equivalence
principle is satisfied. On the other hand, the Einstein equivalence
principle is violated, as there would be a split of photon
helicities\cite{Ni:1975vk}. Furthermore, causality is violated for
timelike $p_{\mu}$. However this violation is significant only in
the regions where the wavelengthes of photons are very large
\cite{causality}.

In summary, current cosmological observations have opened a new
window for probing new physics. In this paper we show that the
current data from WMAP and BOOMERANG might indicate a rotated
polarization angle, which can be resulted from the $CPT$ and Lorentz
violations. Such a result, if confirmed at greater significance by
future observations, would reveal hitherto unknown dynamics of the
nature.

\acknowledgments
We are grateful to T. Montroy  for making the B03 TC and GC
covariance matrix available \cite{TCGC} and kind comments. We thank
G. Hinshaw and L. Page for helpful correspondences. Our fittings
were finished in the Shanghai Supercomputer Center (SSC). We thank
K. Ichiki, A. Lewis, H. Peiris, J. Yokoyama and G. Zhao for helpful
discussions and F. Klinkhamer and R. Lehnert for comments. This work
is supported in part by the National Natural Science Foundation of
China (NSFC) under Grant Nos. 10533010, 90303004, 19925523 and by
the Ministry of Science and Technology of China under Grant No.
NKBRSF G19990754. B. F. is supported by JSPS  and M. L. is by the
Alexander von Humboldt Foundation. X.C. is supported by the China
National Science Fund for Distinguished Scholars (Grant No
10525314).

\end{document}